\documentclass{jfm}

\usepackage{enumitem}

\usepackage{pstool}

\usepackage{color}

\usepackage{amsmath}
\usepackage{amsfonts}
\usepackage{upgreek}

\newcommand{\Smat}{\mathsfbi S}
\newcommand{\Hmat}{\mathsfbi H}

\newcommand{\vect}[1]{\mathbf{ #1 }}

\begin{document}

\shorttitle{Spectral proper orthogonal decomposition} 
\shortauthor{M. Sieber et al.} 

\title{On the nature of spectral proper orthogonal decomposition and related modal decompositions}

\author
 {
  Moritz Sieber
  \corresp{\email{moritz.sieber@fd.tu-berlin.de}},
	C. Oliver Paschereit
  \and 
  Kilian Oberleithner
  }

\affiliation
{
Institut f\"ur Str\"omungsmechanik und Technische Akustik \\
-- Hermann-F\"ottinger-Institut --\\
 M\"uller-Breslau Str. 8, D-10623 Berlin, Germany
}

\maketitle

\begin{abstract}
The spectral proper orthogonal decomposition (SPOD) is a newly introduced extension of snapshot POD that recently gained attention but also brought up controversial issues.
Within the first proposition, the approach was mainly presented in a methodological and phenomenological way.
The present paper will detail the relations between SPOD and related POD approaches from an analytical point of view.
To allow for a better grasp of the approach, an alternative formulation is given that is based on the classic idea from \citet{Lumley.1970} that was carried on by \citet{George.1988}. 
As will be shown, SPOD is closely related to POD with a prior segmentation and Fourier transformation in time.  
Moreover, the SPOD is shown to be equivalent to snapshot POD combined with time delay embedding.
\end{abstract}

\section{Introduction}
The robust identification of coherent structures from experimental or numerical observations of a turbulent flow recently gained interest due to the availability of highly resolved data.
The most promising approach for this task seems to be the proper orthogonal decomposition (POD), which developed from the transfer of statistical methods to the field of fluid dynamics by \citet{Lumley.1970}.
Since than, many variants of the POD developed independently that all aim to improve the results for particular applications and flow configurations.
This independent development of methods, however, lacks some general framework and nomenclature to understand relations between the methods.

In a recent publication \citep{Sieber.2016c}, we proposed an extension of classical snapshot POD, we named spectral proper orthogonal decomposition (SPOD), which overcomes some limitations of the existing approaches.
The SPOD was introduced as a modification of POD that was motivated from data analysis and was proven to be a very useful extension by applying it to experimental data.
From a numerical point of view, the SPOD is just a snapshot POD with an additional filter operation applied to he correlation matrix.
The correlation matrix might be thought of as a condensed representation of all the dynamics of the investigated flow -- a dynamic fingerprint.
The diagonalization of this matrix provides the basis for the snapshot POD \citep{Sirovich.1987}, which comprises the optimum building blocks of the matrix.
The reasoning behind the SPOD approach is that the correlation matrix contains the coherent flow dynamics as well as stochastic fluctuations and measurement noise.
It was observed that the stochastic contributions in the matrix appear as random variations, whereas the coherent structures appear as diagonal streaks in the matrix.
To separate the coherent from stochastic fluctuations, the matrix is filtered such that the random part is attenuated but the coherent dynamics remain unaffected.
It is shown from experimental and numerical data that SPOD provides better assignment of specific fluid dynamic phenomena to individual modes \citep{Sieber.2016,Ribeiro.2017}, in situations where snapshot POD tends to mix several dynamics in single modes. 
Moreover, SPOD compensates for partially recorded phenomena, shows superior noise rejection and provides smooth dynamics that facilitate mode interpretation \citep{Stohr.2017}.

Another prominent POD variant is the \emph{frequency domain} POD \citep{Glauser.1987,Citriniti.2000,Gudmundsson.2011}, which is in some literature also referred to as \emph{spectral} POD \citep{Taira.2017}.
This approach comprises a division of a long time series into short segments, where each segment is Fourier transformed in time and POD is applied to the Fourier coefficients at a specific frequency.
The approach provides a very clean separation of coherent structures in the broad band spectrum of turbulent jets \citep{Gudmundsson.2011}.

In this paper, the SPOD will be put in a more general context in order to reveal its relation to other currently developed POD-based approaches.
We revisit the POD concept of \citet{George.1988,George.2017}, which considers the coupled space and time correlation instead of their separation.
This step back will draw the connections between SPOD and the frequency domain POD outlined in the previous paragraph.
It will be shown that both approaches rely on the same assumptions and that they are the time- and frequency-domain representation of the same decomposition.
Therefore, the two approaches will be named time domain SPOD, when referring to the matrix filter operation and frequency domain SPOD, when referring to the segmentation and Fourier transform in time.

The outline of this paper is the following: 
The first section comprises the derivation of the time domain SPOD from an analytical point of view. 
The classical POD approach is briefly summarized and adapted to detail the new approach.
In this section the focus is put on the time domain SPOD, whereas the next section shows the difference between time- and frequency-domain SPOD.
The subsequent section is dedicated to the similarities of SPOD and time delay embedding.
The latter was recently shown to be a suitable approach to enrich the observation of a nonlinear system such that the system dynamics can be captured by a linear model with intermittent forcing \citep{Brunton.2017}.
The relations between time delay embedding, which is mainly known for dynamic system analysis, and the statistical approach of the SPOD provides a better understanding of botch approaches.

\section{Derivation of SPOD from Lumley's initial POD formulation}

The general concept of POD evolved from the collection of stochastic methods for velocity data by \citet{Lumley.1970}.
According to the retrospective of \citet{George.2017}, POD gained little acceptance in the beginning, despite its obvious capabilities in describing coherent structures of turbulent flows \citep{George.1988}.
The obstacle has always been the collection of sufficient data to obtain converged statistics.

\subsection{The classical POD}

Starting with the decomposition proposed by \citet{George.1988}, we seek the space time correlation function
\begin{align} 
R(x,x',t,t') = \langle u_k(x,t)u_k^*(x',t')\rangle, \label{eq:PODcorr}
\end{align}
where $\langle \rangle$ refers to the ensemble average over repeated observations $()_k$ of a stationary flow $u$ ($^*$ indicates the adjoint).
It is obvious that a statistically relevant number of repeated observations from spatially and temporally resolved data implies a very large amount of data \citep{George.2017}.
From the correlation function $R$, the POD modes $\Phi$ are obtained by solving the \emph{Lumley integral equation}
\begin{align}
\iint{R(x,x',t,t') \Phi_k(x',t')}\mathrm{d}x'\mathrm{d}t' &= \lambda_k \Phi_k(x,t), \label{eq:lumeyint}
\end{align}
which reduces to the commonly known eigenvalue decomposition for the discretized formulation of the problem. 
The obtained POD modes yield the flow decomposition
\begin{align} 
u_j(x,t) = \sum_k \alpha_{jk} \Phi_k(x,t),
\end{align}
where the mode coefficients $\alpha_{jk}$ account for the specific mode amplitude of a certain realization of the flow from the ensemble.
The POD mode shapes $\Phi_k(x,t)$ depend on space and time, which  enables to cover all possible space-time-properties of coherent structures.
In the following, mode shapes and mode coefficients will be abbreviated as modes and coefficients, respectively.
Note that throughout this article integration is always assumed across the entire observed domain and recorded time span.
More details about the treatment of domain boundaries can be found in \citet{Sieber.2016c}.

\subsection{The snapshot POD}
To reduce the necessary amount of data, the time instance at which a \emph{snapshot} of the flow is recorded can be treated as observations of the statistical ensemble.
Therefore, the temporal dependency of the modes is neglected and, in contrast to \eqref{eq:PODcorr}, only the spatial correlation function is considered \citep[p.~68]{Holmes.2012} 
\begin{align} 
R(x,x') = \langle u(x,t_k)u^*(x',t_k)\rangle. \label{POD_spatial_corr}
\end{align}  
A specific property of this approach is the separation of temporal and spatial dependencies of the flow into spatial modes and temporal coefficients
\begin{align} 
u(x,t) = \sum_k a_k(t) \Phi_k(x).
\end{align}
This approach is called snapshot POD as it considered only isolated snapshots and ignores any temporal correlation between them.
This terminology is in contrast to some other publications \citep{Holmes.2012, Taira.2017} where only the computational shortcut via the snapshot correlation (detailed below) is termed \emph{snapshot POD}, but it is in line with the naming used by \citet{George.2017}.
The separation of space and time might contradict the entangled space-time-correlations of turbulent flows; however, it provides a mathematical rigor that tremendously simplifies the governing equations employing the Galerkin projection \citep{Noack.2013}.
Extensive description of possible reduced order modeling and detailed properties of the snapshot POD are collected in \citet{Holmes.2012}.
Moreover, the numerical implementation of snapshot POD allows for the interchange of spatial correlation with a snapshot correlation that leads to a much faster computation for typical flow data \citep{Sirovich.1987}.

As shown by \citet{Aubry.1991}, the interchange between spatial and temporal correlation is not only possible in the discrete but also in the continuous case, given that the data in spatial and temporal domain belong to the space of square integrable functions.
Accordingly, the solution of integral equation $\int{C(t,t')a(t')\mathrm{d}t'} = \lambda a(t)$ with the snapshot correlation function
\begin{align} 
C(t,t') = \int{u(x,t)u^*(x,t')}\mathrm{d}x \label{POD_snap_corr}
\end{align} 
provides the coefficients $a(t)$ that likewise provide the modes $\Phi(x)$ by expanding the temporal evolution of the data by the coefficients (see also \eqref{eq:SPODmode}).
The perfect symmetry between the temporal and spatial expansion inspired \citet{Aubry.1991} to name this approach bi-orthogonal decomposition.

\subsection{The spectral POD (SPOD)}

The following section gives an alternative formulation to snapshot POD that accounts for the temporal evolution of the flow without the need for vast amounts of data.
The approach adopts the segmentation of time series that is used in Welch's method to estimate the power spectral density from limited data.  
The approach is commonly used for spectral averaging and was also applied to POD analysis (e.g. \cite{Citriniti.2000,Taira.2017}).
However, in contrast to these applications no Fourier transform of the segments is employed here.
To include this methodology in the present derivation, we use a temporal weighting (window) function that is multiplied with the time series to isolate a segment.

The derivation is started from the previous formulation \eqref{eq:PODcorr} but we seek to compute the temporal correlation as well as ensembles from a long time series.
Therefore, we introduce a short time scale $\tau$ across which the flow is correlated and a window function $w(\tau)$ that accounts for the observation horizon $T$.
The window function is taken to be a Gaussian function $w(\tau) = \mathrm{exp}(-(\tau/T)^2)$.
The time series is segmented into windows centered at discrete times $t_k$ constituting the observed ensembles.
Consecutive segments may also overlap to an arbitrary extend. 
By inserting the two time scales into \eqref{eq:PODcorr}, the space-time-correlation function reads
\begin{align}
R(x,x',\tau,\tau') &= \langle u(x,t_k+\tau)w(\tau)u^*(x',t_k+\tau')w^*(\tau') \rangle \label{SPODcorr}\\
 &=  \int u(x,t+\tau)w(\tau)u^*(x',t+\tau')w^*(\tau') \mathrm{d}t,\label{SPODcorr2}
\end{align}
where in the second expression the ensemble average is replaced by a time average.
The change from discrete ensembles to continuous integration corresponds to the maximal overlap of consecutive time segments.
Note that the increase of the overlap and consequent addition of correlated ensembles does not change the resulting POD modes.
The assumption of uncorrelated ensembles (snapshots) as noted in various related literature is only due to the to minimization of the computational costs \citep{Sirovich.1987}.

Equivalent to \eqref{eq:lumeyint} the POD modes are derived from the integral equation 
\begin{align}
\iint{R(x,x',\tau,\tau') \Phi_k(x',\tau')}\mathrm{d}x'\mathrm{d}\tau' &= \lambda_k \Phi_k(x,\tau) \label{eq:lumeyint_tau}.
\end{align}
The resulting modes $\Phi(x,\tau)$ are not only functions of space but also represent the temporal evolution across a time window given by $w(\tau)$.

Following the argumentation of \citet{Aubry.1991}, we may compute the POD coefficients from the snapshot correlation, instead of the modes from space-time-correlation. 
Therefore, the integration variable of \eqref{SPODcorr2} is changed from $t$ to $x \times \tau$, and the correlation among different times $t$ and $t'$ is investigated. 
This interchange is analogously to the step from \eqref{POD_spatial_corr} to \eqref{POD_snap_corr} of the snapshot POD, with the difference that for the SPOD formulation the additional integration across $\tau$ adds to the snapshot correlation function
\begin{align}
S(t,t') =& \iint{ u(x,t+\tau)w(\tau)u^*(x,t'+\tau)w^*(\tau) } \mathrm{d}x \mathrm{d}\tau \label{SPODmatrix1}\\
 =& \int{ w^2(\tau) C(t+\tau,t'+\tau)  } \mathrm{d}\tau  \label{SPODmatrix}.
\end{align}
The simplification from \eqref{SPODmatrix1} to \eqref{SPODmatrix} is due to the proposed Gaussian weighting and use of \eqref{POD_snap_corr}.
Writing the equation for discrete time instances $t = k \Delta t$, $t' = l \Delta t$ and $\tau = j \Delta t$ using appropriate quadrature, the correlation matrix reads
\begin{align}
S_{k,l} = \sum_{j=-N_f}^{N_f}{g_j C_{k+j,l+j}} \label{eq:SPODfilt}.
\end{align}
This expression is the filtered version of the snapshot correlation matrix that was proposed in our previous paper \citep{Sieber.2016c}.
This analogy implies that the coefficients $g_j = w^2(j \Delta t)$ are the previously introduced SPOD filter coefficients that specify a discrete filter of extend $1+2N_f$. 
Hence, the SPOD concept outlined here is a generalization of the approach in \citet{Sieber.2016c}.

After this short detour to the discrete variables (more can be found in \citet{Sieber.2016c}) we proceed with the continuous formulation of the snapshot correlation matrix given in \eqref{SPODmatrix}.
The corresponding SPOD coefficients are determined by solving
\begin{align}
\int S(t,t') a_k(t') \mathrm{d}t' = \lambda_k a_k(t) \label{eq:SPODeigenvalue}.
\end{align}
The coefficients $a_k(t)$ form an orthogonal basis in time domain and are scaled to fulfill $\int{a_k(t) a_l(t)}\mathrm{d}t = \lambda_k \delta_{kl}$.
The SPOD modes that also extend in $\tau$ direction are determined from the convolution
\begin{align}
\Phi_k(x,\tau) = \frac{1}{\lambda_k}\int{a_k(t) u(x,t+\tau) w(\tau)} \mathrm{d}t \label{eq:SPODmode}
\end{align}
and are likewise orthonormal with $\iint{\Phi_k(x,\tau) \Phi^*_l(x,\tau)}\mathrm{d}x\mathrm{d}\tau = \delta_{kl}$.
Finally, for the decomposition of the original flow data only the mode centered at $\tau = 0$ is required, reading
\begin{align}
u(x,t) = \sum_{k=1}^{N}{a_k(t) \Phi_k(x,\tau=0)} \label{eq:SPODdecomp}.
\end{align}
These partial modes are not necessarily orthogonal \citep{Sieber.2016c}, but they are based on an orthogonal spatio-temporal mode base that has a compact support in time as defined by the temporal window function $w(\tau)$. 

Similar to \eqref{eq:SPODmode}, the SPOD coefficients are derived from the modes by the convolution
\begin{align}
a_k(t) = \iint{\Phi_k(x,\tau) u^*(x,t+\tau) w^*(\tau)} \mathrm{d}x \mathrm{d}\tau \label{eq:SPODcoeff},
\end{align}
which is in contrast to the snapshot POD coefficients, where $a_k(t) = \int{\Phi_k(x) u^*(x,t)} \mathrm{d}x$.
The comparison of the two expressions shows the essential differences between POD and SPOD. 
While the coefficients of the POD only reflect the temporal correlation that is present in the data at a certain time, the SPOD involves a temporal convolution across the short time $\tau$ and, therefore, maintains the temporal continuity.

\section{Time vs. frequency domain SPOD}

The previous section describes the time domain SPOD that is equivalent to the formulation proposed in \citet{Sieber.2016c}. 
In this section we show its relation to the frequency domain SPOD that has been applied in a number of previous studies \citep{Glauser.1987,Citriniti.2000,Gudmundsson.2011}.
As will be shown, the methodology of frequency domain SPOD is largely the same as for the time domain SPOD.
It likewise involves the segmentation and weighting of a long time series with the additional step of a Fourier transformation in time $\tau$ \citep{Taira.2017}.
Analogously to the derivation of \eqref{SPODcorr2}, the spatial correlation function of the frequency domain SPOD can be expressed as
\begin{align}
R(x,x',\omega) &=  \int \int{u(x,t+\tau)w(\tau)e^{-i\omega\tau}}\mathrm{d}\tau \int{u^*(x',t+\tau')w^*(\tau')e^{i\omega\tau'}}\mathrm{d}\tau' \mathrm{d}t \label{corr_SPODfreq}\\
& = \iint{R(x,x',\tau,\tau')e^{i\omega(\tau'-\tau)}}\mathrm{d}\tau\mathrm{d}\tau', \label{corr_trafo}
\end{align}
where $i$ indicates the imaginary unit and the integrals are rearranged from \eqref{corr_SPODfreq} to \eqref{corr_trafo} such that the correlation function \eqref{SPODcorr2} is isolated.
Accordingly, the frequency domain correlation function is either given by the correlation of the Fourier transformed velocity signal \eqref{corr_SPODfreq} or by the Fourier transform of the time-domain correlation function \eqref{corr_trafo}. 
The corresponding frequency-depended modes $\Phi_k(x,\omega)$ can be obtained from the integral equation
\begin{align}
 \int{R(x,x',\omega)\Phi_k(x',\omega)\mathrm{d}x'} = \lambda_k \Phi_k(x,\omega). \label{inteq_freq}
\end{align}

The relation between the modes from time domain SPOD and frequency domain SPOD is revealed by the following derivation.
We start with the time domain integral equation \eqref{eq:lumeyint_tau}, which is multiplied with $e^{-i\omega\tau}$ and integrated across $\tau$ ($\int \mathrm{d}\tau$), giving
\begin{align}
\iiint{R(x,x',\tau,\tau') \Phi_k(x',\tau') e^{-i\omega\tau}} \mathrm{d}\tau \mathrm{d}\tau'\mathrm{d}x' = \lambda_k \int{ \Phi_k(x,\tau) e^{-i\omega\tau} \mathrm{d}\tau}. \label{deriv1}
\end{align}
This equation is simplified assuming a statistically stationary time series, which gives a correlation function that only depends on time differences, giving $R(x,x',\tau,\tau') = R(x,x',\tau'-\tau)$ \citep{George.1988}. 
Rearranging \eqref{deriv1} and inserting  \eqref{corr_trafo} ultimately leads to
\begin{align}
\int{R(x,x',\omega) \int{\Phi_k(x',\tau') e^{-i\omega\tau'}} \mathrm{d}\tau' } \mathrm{d}x' = \lambda_k \int{ \Phi_k(x,\tau) e^{-i\omega\tau} \mathrm{d}\tau}. \label{deriv2}
\end{align}
The comparison of \eqref{deriv2} with \eqref{inteq_freq} leads to the conclusion that the modes that result from the frequency domain SPOD are identical with the Fourier transform of the time domain modes that reads as
\begin{align}
\Phi_k(x,\omega) = \int{\Phi_k(x,\tau)e^{-i\omega\tau}\mathrm{d}\tau} \label{mode_trafo}.
\end{align}

Analogously to the conversion from spatial \eqref{SPODcorr2} to snapshot correlation \eqref{SPODmatrix1} for the time domain SPOD, the frequency domain SPOD can also be deduced from the coefficients instead of the modes.
By changing the integration of \eqref{corr_SPODfreq} from time to space and seeking the cross-correlation between $t$ and $t'$, the frequency domain snapshot correlation function is obtained from
\begin{align}
S(t,t',\omega) &=  \int \int{u(x,t+\tau)w(\tau)e^{i\omega\tau}}\mathrm{d}\tau \int{u^*(x,t'+\tau')w^*(\tau')e^{-i\omega\tau'}}\mathrm{d}\tau' \mathrm{d}x \label{eq:freq_SPOD_snapshot1}\\
 &=  \iint C(t+\tau,t'+\tau') w(\tau)w^*(\tau') e^{i\omega(\tau-\tau')} \mathrm{d}\tau \mathrm{d}\tau' \label{eq:freq_SPOD_snapshot2},
\end{align}
where the integrals are rearranged and the snapshot correlation function $C$ is isolated as done in \eqref{SPODmatrix}.
Equations \eqref{eq:freq_SPOD_snapshot1} and \eqref{eq:freq_SPOD_snapshot2} show that, similar to the spatial correlation function \eqref{corr_SPODfreq} and \eqref{corr_trafo}, the snapshot correlation function in frequency domain can either be obtained from the correlation of the Fourier transformed velocity \eqref{eq:freq_SPOD_snapshot1} or the Fourier transform of the POD snapshot correlation function \eqref{eq:freq_SPOD_snapshot2}.
However, the Fourier decomposition of the POD snapshot correlation function $C$ is not trivial since it involves a transform in direction ($\tau-\tau'$) and subsequent average of the Fourier coefficients in ($\tau+\tau'$)  direction. 
The averaging of correlation function $C$ in ($\tau+\tau'$) direction is also pursued for the time domain SPOD \eqref{SPODmatrix} and can therefore be factored out (for Gaussian weighting $w$) to simplify \eqref{eq:freq_SPOD_snapshot2} to
\begin{align}
S(t,t',\omega) &= \int{ S(t+\tau/2,t'-\tau/2) w(\tau) e^{-i\omega\tau} }\mathrm{d}\tau.
\end{align}
The inversion of this transformation is given by 
\begin{align}
S(t,t') &=  \int S(t,t',\omega) \mathrm{d}\omega,
\end{align}
which allows for the interchange between the time- and frequency-domain SPOD for the snapshot correlation.

The fact that both SPOD approaches can be converted into each other and rely on similar assumptions affirmed the choice to name both SPOD.
The two variants are distinguished by the treatment of the temporal dimension, and therefore, it is either a time domain SPOD or a frequency domain SPOD.

\section{Relation of time-delay embedding and SPOD}

Time-delay embedding is a method to enrich the information that can be drawn from limited observations of a system.
It builds on \citet{Takens.1981} embedding theorem, which states that the state of a continuous chaotic attractor can be reconstructed from a single measurement time-series.
Accordingly, a single measurement is enriched with time-delayed measurements to form a larger basis.
Delay embedding has shown to improve flow control \citep{Brunton.2015}, stochastic estimation \citep{Lasagna.2013} and reduced order modeling \citep{Brunton.2017}, to name just a few examples related to fluid dynamics.

In the following we show that there is a very close relation between the delay embedding and the short time-scale $\tau$ introduced above.
For this purpose, we leave the continuous framework of the previous sections and consider the collection of $P$ measured points taken at $N$ instants of time, represented by the vector $\vect{u}(t) = [u(x_1,t), u(x_2,t), \cdots , u(x_P,t)]^T$.
Furthermore, $M$ additional delay coordinates are created by shifting the sequence in time.
The data are collected in a block-Hankel matrix (referring to the nomenclature of \citet{Brunton.2017})
\begin{align}
\Hmat = \left[ 
\begin{array}{cccc}
\vect{u}(t_1) & \vect{u}(t_2) & \cdots & \vect{u}(t_{N-M})\\
\vect{u}(t_2) & \vect{u}(t_3) & \cdots & \vect{u}(t_{N-M+1})\\
\vdots & \vdots & \ddots & \vdots\\ 
\vect{u}(t_M) & \vect{u}(t_{M+1}) & \cdots & \vect{u}(t_{N})
\end{array}
\right],
\end{align}
which is regarded as a snapshot matrix and decomposed with POD.
The corresponding snapshot correlation matrix is $\Smat = \Hmat^T\Hmat$, which can likewise be describe by the element-wise summation
\begin{align}
S_{k,l} = \sum_{j=1}^M\sum_{p=1}^P{u(x_p,t_{k+j})u(x_p,t_{l+j})}. \label{eq:delay_snap_corr}
\end{align}
With the definition of the POD correlation matrix as $C_{k,l} = \sum_{p=1}^P{u(x_p,t_k)u(x_p,t_l)}$, the RHS of \eqref{eq:delay_snap_corr} can be expressed as 
\begin{align}
S_{k,l} = \sum_{j=1}^M{C_{k+j,l+j}},
\end{align}
which reveals that the time delay embedding is equivalent to a summation of the diagonals of $C_{k,l}$. 
Comparing this result to \eqref{eq:SPODfilt} shows that POD in combination with time delay embedding is equivalent to SPOD with a box filter ($g_j = 1$ and $N_f = {(M-1)}/{2}$).

In addition to this correspondence, the SPOD has further unique characteristics.
In contrast to time delay embedding, the SPOD includes a weighting of the delay with respect to the amount of delay relative to the center coordinate.
Moreover, delay embedding requires the construction of a very large Hankel matrix, while for SPOD the construction of this matrix is implicitly assumed but never actually performed. 
Instead, the effect of the time delay embedding is replicated by the filter applied to the correlation matrix.
This saves computation time when the number of snapshots is smaller than the number of spatial points times the embedding dimension ($N<PM$).

\section{Conclusions}
\begin{figure}
	\centering
		\includegraphics[width=0.99\textwidth]{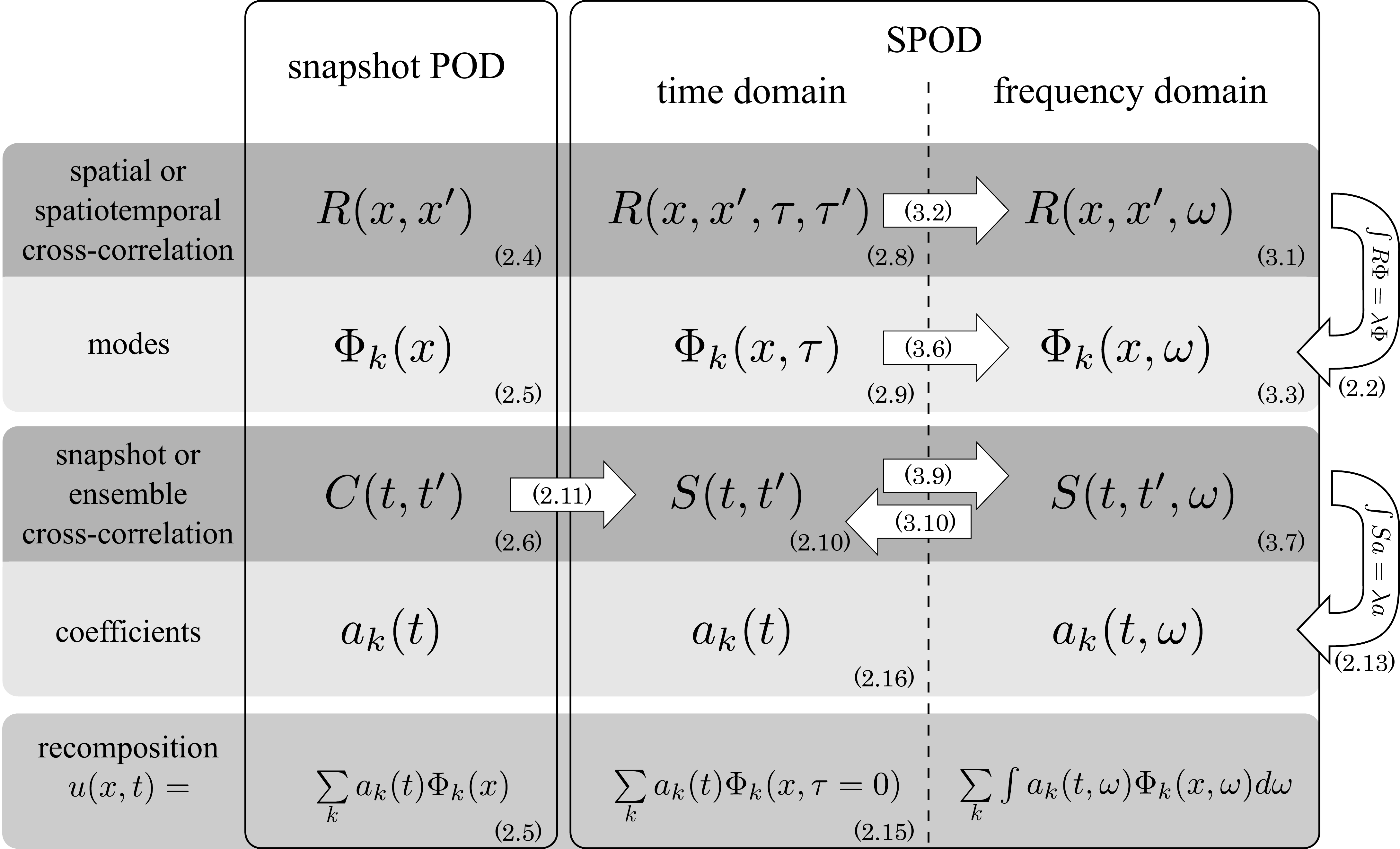}
	\caption{Classification of different POD approaches by the dependencies of the functions on the governing variables. The numbers in brackets indicate the corresponding equations in the paper. The horizontal arrows indicate possible conversion of the functions.}
	\label{fig:methods}
\end{figure}

In this paper we present an analytical framework that links several concurrent POD approaches.
Figure \ref{fig:methods} shows a graphical representation of these different POD variants. 
Each column refers to a different POD approach, while the rows list the corresponding correlation functions, modes and coefficients.

As can be seen in the upper rows of figure \ref{fig:methods}, the fundamental difference between snapshot POD and SPOD is the treatment of the space-time correlation. 
For the snapshot POD, spatial structure and temporal evolution are separated in a product approach, which may lead to unphysical modes.
The SPOD, instead, considers the spatio-temporal evolution of the modes across a short time span.
The differences between snapshot POD and SPOD can be understood as such that snapshot POD reveals statistically similar states among all recorded instances, whereas SPOD reveals similar space-time trajectories within the recorded data. 
The snapshot correlation functions that are sketched in the mid row of figure \ref{fig:methods} do not exhibit these differences in such a clear way.
Nevertheless, their representation highlights the close relation among the different approaches and outlines numerical shortcuts to compute the SPOD trough simple manipulations of the snapshot correlation functions.

Despite their similarities, there are conceptual differences between the time- and frequency-domain SPOD, which are of relevance for specific applications.
The re-composition of the flow from the modes and coefficients given in the bottom row of figure \ref{fig:methods} shows that the time domain SPOD allows for a re-composition that is very similar to the snapshot POD, while the frequency domain SPOD requires an additional integration along all frequencies.
Therefore, time domain SPOD is more suitable for reduced order modeling, however, keeping in mind that the reduced modes ($\tau=0$) not necessarily form a orthogonal basis.
The frequency domain SPOD separates the flow into pure harmonic contributions that have trivial dynamics but give additional constraints that are beneficial to characterize broad band dynamics.

Interestingly, in the relevant literature the frequency domain SPOD appears to be mainly applied to flows with convective instabilities (amplifier type flow, e.g. free jets, boundary layers), whereas the snapshot POD and time domain SPOD  seem to be favorably applied to flows with global instabilities (oscillator type flow, e.g. cylinder wake, cavity flow).
This distinctive selection of methods relates to the prior expectations about the properties of the modal decomposition.
The options are either the spectral purity provided by the frequency domain SPOD or the modal sparsity (optimality) provided by the time domain SPOD and snapshot POD.
For amplifier type flows a dispersed spectrum of modes is expected and therefore the spectral purity is preferred.
Instead, an oscillator flow is expected to have sparse dynamics, which are desired to be captured by the decomposition with as few modes as possible.
The better understanding of the two SPOD variants give the freedom to fade between the strict classifications and therefore may help to reveal new flow structures.

The close similarity between delay embedding and time domain SPOD explains the previously reported capability of SPOD to reconstruct dynamics of partially recorded phenomena \citep{Sieber.2016c}, which is the main feature of delay embedding.
Moreover, it explains why delay embedding in combination with SVD (same as SPOD) by \citet{Brunton.2017} tends to give coefficients that can be described by linear models.
This is seen from the fact that SPOD might be understood as a local linearization of the dynamics by restricting the change of frequency and amplitude of the modes \citep{Sieber.2016c}.
Similarly, the HAVOK models \citep{Brunton.2017} tend to exhibit linear dynamics and the nonlinearities contribute as intermittent forcing.

\subsection{Acknowledgment}\label{acknowledgment}
The authors kindly acknowledge the funding from the German Research Foundation (DFG Project PA 920/30-1, DFG Project PA 920/37-1) and from the Research Association for Combustion Engines (FVV). 

\bibliographystyle{jfm}
\bibliography{references}

\end{document}